\documentstyle[11pt,aaspp4]{article}
\newcommand{\onu}{\Omega_\nu}
\newcommand{\oc}{\Omega_c}
\newcommand{\om}{\Omega_{\rm m}}
\newcommand{\ov}{\Omega_\Lambda}
\newcommand{\ob}{\Omega_{\rm b}}

\def\go{\mathrel{\raise.3ex\hbox{$>$}\mkern-14mu
\lower0.6ex\hbox{$\sim$}}}
\def\lo{\mathrel{\raise.3ex\hbox{$<$}\mkern-14mu
\lower0.6ex\hbox{$\sim$}}} \def\onu{\Omega_\nu}

\begin{document}
\title{Redshift Evolution of the Nonlinear Two-Point Correlation
Function} 
\author{Chung--Pei Ma} 
\affil{Department of Physics and
Astronomy, University of Pennsylvania, Philadelphia, PA~19104;
cpma@strad.physics.upenn.edu}

\begin{abstract}  
This paper presents a detailed theoretical study of the two-point
correlation function $\xi$ for both dark matter halos and the matter
density field in five cosmological models with varying matter density
$\om$ and neutrino fraction $\onu$.  The objectives of this systematic
study are to evaluate the nonlinear gravitational effects on $\xi$, to
contrast the behavior of $\xi$ for halos vs. matter, and to quantify
the redshift evolution of $\xi$ and its dependence on cosmological
parameters.  Overall, $\xi$ for halos exhibits markedly slower
evolution than $\xi$ for matter, and its redshift dependence is much
more intricate than the single power-law parameterization used in the
literature.  Of particular interest is that the redshift evolution of
the halo-halo correlation length $r_0$ depends strongly on $\om$ and
$\onu$, being slower in models with lower $\om$ or higher $\onu$.
Measurements of $\xi$ to higher redshifts can therefore be a potential
discriminator of cosmological parameters.  The evolution rate of $r_0$
for halos within a given model increases with time, passing the phase
of fixed comoving clustering at $z\sim 1$ to 3 toward the regime of
stable clustering at $z\sim 0$.  The shape of the halo-halo $\xi$, on
the other hand, is well approximated by a power law with slope $-1.8$
in all models and is not a sensitive model discriminator.
\end{abstract}
\centerline{To appear in the {\it Astrophysical Journal} (Vol. 510, 
January 1 1999)}

\section{Introduction}
The two-point correlation function $\xi(r)$ is a fundamental quantity
in cosmology.  For the density field of matter, it provides the most
basic statistical measure of gravitational clustering, and its Fourier
pair is simply the power spectrum $P(k)$.  For galaxies, $\xi$ offers
a convenient measure of their clustering strength, and a determination
of $\xi$ is among the most important goals of every major galaxy
survey.  Although $\xi$ is not independent from $P$, the estimators
for the two quantities have different error properties and therefore
both carry useful information.

The two-point correlation function of nearby galaxies in physical
space has been determined in several galaxy redshift surveys and is
found to be well approximated by a power law, $\xi=(r/r_0)^{-\gamma}$,
over the distance range $1 < r < 20\,h^{-1}$ Mpc.  Analysis of the
early CfA sample of 2400 galaxies (complete to $m_B=14.5$) gives a
correlation length of $r_0 = 5.4 \pm 0.3\,h^{-1}$ Mpc and a slope of
$\gamma=1.77$ for $1 \lo r \lo 10\,h^{-1}$ Mpc (Davis \& Peebles
1983).  Subsequent optical surveys generally have yielded consistent
results.  The combined CfA2 and Southern Sky Redshift Surveys of 12812
galaxies, for example, give $r_0 = 4.95 \pm 0.13\,h^{-1}$ Mpc and
$\gamma=1.73\pm 0.03$ when cluster galaxies are removed (Marzke et
al. 1995).  The Las Campanas Redshift Survey (Shectman et al. 1996) of
19558 galaxies is found to have $r_0= 5.06 \pm 0.12\,h^{-1}$ Mpc and
$\gamma=1.862\pm 0.034$ (Jing, Mo, and Borner 1998).  Surveys at other
wavelengths have shown larger variations.  The 1.2 Jy
IRAS survey of 10000 galaxies yields $r_0 = 3.76\,h^{-1}$ Mpc and
$\gamma=1.66$ for $r< 20\,h^{-1}$ Mpc (Fisher et al. 1994).  A small
sample of 300 1.4 GHz radio sources is found to have $r_0= 11\,h^{-1}$
Mpc and $\gamma=1.8$, where the enhanced correlations are attributed
to the high optical luminosity and rich environments inhabited by
radio galaxies (Peacock \& Nicholson 1991).

Although measurements of $\xi(r)$ for nearby galaxies are abundant,
the two-point correlation function at higher redshifts is an important
quantity that has only begun to be determined.  Deep redshift surveys
are clearly required for this purpose, and initial results have been
somewhat conflicting.  The Autofib survey of 1100 galaxies from 33
pencil beams, for example, shows no evidence of evolution, giving
$r_0=6.2 \pm 0.4\,h^{-1}$ Mpc and $r_0=6.8 \pm 0.4\,h^{-1}$ Mpc for
the samples below and above the survey median redshift $\langle z
\rangle=0.16$, respectively (Cole et al. 1994).  The Canada-France
Redshift Survey, on the other hand, finds a decreasing correlation
function with increasing redshift, with $r_0=1.33 \pm 0.09\,h^{-1}$
Mpc at $z=0.53$ (Le Fevre et al. 1996).  A sample of 183 field
galaxies covering 216 square arcmin from the CNOC (Canadian Network
for Observational Cosmology) cluster survey yields $r_0=2.2\pm
0.5\,h^{-1}$ Mpc for $\langle z \rangle=0.37$ (Shepherd et al. 1997).
A sample of 248 K-band selected galaxies from two fields of combined
area of 27 square arcmin gives $r_0=2.9\pm 0.34\,h^{-1}$ Mpc at
$z=0.34$, $r_0=2.0\pm 0.3\,h^{-1}$ Mpc at $z=0.62$, and $r_0=1.4\pm
0.21\,h^{-1}$ Mpc at $z=0.97$ (Carlberg et al. 1997).  The shape of
$\xi$ shows little variation with redshift and is well fit by a power
law with a slope $\gamma = 1.7$ to 1.8 in all surveys.  More
measurements from deep surveys covering larger areas of the sky will
provide more robust determination of $\xi(r)$ beyond $z\approx 0$.

On the theoretical side, models for the redshift evolution of $\xi(r)$
have been simplistic for the most part, and the interpretation of
observational results is often cast either in the regime of stable
clustering in physical space ($\xi\propto a^3$, where $a=(1+z)^{-1}$),
or fixed clustering in comoving space.  Uncertainties and statistical
fluctuations in older data sets due to small galaxy numbers and
limited survey fields are often too large to allow a finer distinction
beyond these two regimes, but data from the ongoing and planned future
deep redshift surveys certainly have the potential for discriminating
among cosmological parameters, if the evolution of $\xi(r)$ follows
distinct patterns in different models.  This provides the motivation
for the extensive investigation of $\xi$ presented here for a wide
range of redshift and cosmological models.

Theoretical predictions of $\xi$ fall in two categories: the density
field and dark matter halos.  Much work has been done for the former
(Jain 1997; Sheth and Jain 1997; Matarrese et al. 1997; Jenkins et
al. 1998; and references therein), particularly in the context of
scale-free models with power-law $P(k)$ and the cold dark matter (CDM)
models.  No systematic study, however, has been carried out to
investigate a spectrum of models with different matter densities as
well as neutrino masses over a large redshift range.  Theoretical
calculations of $\xi$ for the density field can enhance our general
understanding of gravitational clustering, but the results cannot be
compared directly with the observed $\xi$ without the knowledge of the
relationship between the density field and galaxies.

The two-point correlation function of dark matter halos, on the other
hand, bears closer resemblance to the observed $\xi$.  Analytical bias
models have been constructed to relate $\xi$ for the density field to
halos (Moscardini et al. 1998), but a more direct, less
model-dependent approach is to compute the halo-halo $\xi$ from
numerical simulations of structure formation.  Since this requires
more extensive effort, only a few individual models have been studied.
Gelb \& Bertschinger (1994b), Brainerd \& Villumsen (1994), Bagla
(1997), and Jing (1998), for example, examined $\xi$ of both matter
and halos in the standard CDM model (although with different halo
identification schemes.  See \S~3.2).  The correlation function in the
$\onu=0.3$ cold+hot dark matter (C+HDM) model has been studied in
Klypin et al. (1993) and Jing et al. (1994), but this model has been
shown to possess too little small-scale power at high redshifts to
account for the amount of observed collapsed objects.  Colin et
al. (1997) is the only work thus far that has specifically quantified
the redshift evolution of halo-halo $\xi$ in different models.  Both
$\om=1$ and $\om=0.2$ CDM models were examined, but their choice of
high Hubble constant, $h=1$, exceeds all current observational values
and also gives the $\om=1$ CDM power spectrum a shape parameter of
$\Gamma=1$, which is far above $\Gamma\approx 0.25$ indicated by
galaxy clustering and the value of $\Gamma=0.5$ for the standard CDM
model.

In this paper I present results from an extensive theoretical study of
the properties and evolution of $\xi(r)$ in three classes of structure
formation models that are of current cosmological interest: CDM,
C+HDM, and flat low-density CDM models with a cosmological constant
(LCDM).  The objectives are to (1) evaluate the nonlinear
gravitational effects on $\xi$; (2) compare the behavior of $\xi$ for
matter vs. halos, which reflects the bias between matter and halos and
is sometimes a confused issue in the literature; and (3) quantify the
redshift evolution of $\xi$ and its dependence on cosmological
parameters.  No such systematic study over this wide range of
cosmological models and redshift has been carried out thus far.  In
order to determine the nonlinear evolution of $\xi$, numerical
simulations of structure formation in five flat models with matter
density $\om=1$, 0.5, 0.3, and neutrino fraction $\onu=0.1$ and 0.2
are performed (see \S~2).  Results of $\xi$ for both the density field
and halos from linear as well as nonlinear calculations are discussed
in \S~3.  Particular emphasis is placed on quantifying and contrasting
the redshift dependence of the halo-halo $\xi$ in different
cosmogonies, which can serve as a potential discriminator of
cosmological parameters when more high-redshift data are collected.

\section{Simulations of Nonlinear Clustering}
Numerical simulations are performed to determine the nonlinear
two-point correlation functions for both the density field and dark
matter halos in five cosmological models that include two C+HDM, two
LCDM, and the standard CDM model.  For this study to form a controlled
numerical experiment, the simulations for the C+HDM and LCDM models
reported here are all performed in a (100 Mpc)$^3$ comoving box with a
Plummer force softening length of 50 kpc comoving (for the given value
of the Hubble constant in each model), and identical phases are used in
the initial conditions for the runs.  Two additional large simulations
in a (640 Mpc)$^3$ box with a force softening length of 160 kpc and a
different set of initial conditions are used to test the effects of
numerical resolution and realization on our results (see \S~3).  The
primordial power spectra for all models have a spectral index of
$n=1$, and the density fluctuations are drawn from a random Gaussian
distribution.  The gravitational forces are computed with a
particle-particle particle-mesh (P$^3$M) code (Bertschinger \& Gelb
1991; Gelb \& Bertschinger 1994ab; Ma \& Bertschinger 1994b).  The
number of simulation particles used to represent the cold dark matter
is $256^3$ for the 640 Mpc CDM run, and $128^3$ for the rest.  For the
hot dark matter component in C+HDM models, $128^3$ and $10\times
128^3$ particles are used in the $\onu=0.1$ and 0.2 models,
respectively.  Although the latter is needed for a fine sampling of
the neutrino velocity phase space (Ma \& Bertschinger 1994a), the
two-point correlation function -- the subject of study here -- mostly
probes the spatial distribution of matter.  Tests with a reduced
number of hot particles have been performed in the $\onu=0.2$ model to
verify that $128^3$ is sufficient for an accurate determination of
$\xi$.

For the C+HDM models, only $\onu=0.1$ and $\onu=0.2$ are studied here
because structure forms too late in models with $\onu > 0.2$ to
account for high-redshift galactic structures (Ma et al. 1997 and
references therein).  A baryon fraction of $\ob=0.05$ and a Hubble
constant of $h=0.5$ are assumed, and the cold dark matter fraction is
$\Omega_{\rm cdm}=0.85$ and 0.75 for the two models, respectively.
The rms linear mass fluctuation in spheres of radius $8\,h^{-1}$ Mpc
is $\sigma_8=0.9$ and 0.82 for $\onu=0.1$ and 0.2, when the 4-year
COBE result (Bennett et al. 1996; Gorski et al. 1996) is used as the
normalization.

The two LCDM models chosen for the simulations have
$(\om,\ov,h)=(0.3,0.7,0.75)$ and $(0.5,0.5,0.7)$.  The parameter
$\Gamma=\om\,h$ characterizing the shape of the power spectrum is 0.23
and 0.35 for the two models.  Both models are COBE-normalized, with
$\sigma_8=1.28$ and 1.53, respectively.  The COBE-normalized standard
CDM model with $\om=1$ and $h=0.5$ (no baryons) is also included in
this study as a point of standard reference despite of its excess
power on small scales.  The corresponding $\sigma_8=1.4$ is a factor
of $\sim 2$ higher than that required by cluster abundance, but the
results quoted here can be easily translated to a lower normalization
by rescaling the redshift.  The results for the standard CDM model are
computed from a P$^3$M simulation performed by Gelb \& Bertschinger
(1994ab), which employed a numerical algorithm and numerical
parameters similar to the LCDM and C+HDM runs used here: 66 kpc for
the Plummer force softening distance, 100 Mpc box, and $144^3$
particles.  The baryon fraction is set to zero for both LCDM and the
standard CDM models so that the effects of changing $\om$ can be
clearly studied.

\section{Two-Point Correlation Function}
The two-point correlation functions $\xi$ for the mass density field
and dark matter halos have different shapes and follow distinct
evolutionary histories.  They are discussed separately in \S~3.1 and
\S~3.2.  Results for the evolution of the correlation length are
discussed in \S~3.3.  All figures for $\xi$ are presented in physical
coordinates $r$, but its behavior in comoving space is also discussed.

\subsection{$\xi$ of the Density Field}
The two-point correlation function of the density field measures the
lowest-order gravitational clustering of matter in a given
cosmological model.  It is the Fourier transform of the power spectrum
$P(k)$ for the density fluctuation $\delta\rho/\rho$, and can be
written as
\begin{equation}
 \xi(r)=\int_0^\infty {dk\over k}\,{4\pi k^3\,P(k)} {\sin{kr}\over kr}\,.
\label{xi}
\end{equation}
In the linear regime, $\xi$ is easily calculated from the linear
$P(k)$, which can be obtained from either direct integration of the
coupled Boltzmann and Einstein equations, or analytical approximations
for various cosmological models that have been published in the
literature (e.g., Bardeen et al. 1986; Efstathiou et al. 1992; Ma
1996).  In the nonlinear regime, $\xi$ can be calculated either from
the particle positions in numerical simulations directly, or from
integration of the few published fitting functions for the nonlinear
$P(k)$.  For the nonlinear $P(k)$, Hamilton et al. (1991) examined
scale-invariant models with power-law spectra, Jain et al. (1995) and
Peacock \& Dodds (1996) studied the flat and open CDM models, while Ma
(1998) provided a general approximation valid for CDM, LCDM, as well
as C+HDM models.

Figures~1 to 3 compare $\xi(r)$ for the linear and nonlinear density
fields at various redshifts for the five models chosen for this study
-- the standard CDM (Fig.~1), the $\om=0.5$ and 0.3 LCDM models
(Fig.~2), and the $\onu=0.1$ and 0.2 C+HDM models (Fig.~3).  The
dashed curves show the linear $\xi$ computed from equation~(\ref{xi}).
For the CDM and LCDM models, the input linear $P(k)$ is the Bardeen et
al. fitting function (1986) with the shape parameter $\Gamma=\om\,h$
(no baryons are assumed in these models).  For the C+HDM models, the
input is the fitting function of Ma (1996) for the density-averaged
linear $P(k)$, where $P(k)=[\onu\,P_\nu^{1/2}+ (1-\onu)\,P_c^{1/2}]^2$
measures the total gravitational fluctuations contributed by the
separate hot and cold components.

The circles in Figures~1 to 3 show the nonlinear $\xi$ calculated
directly from the particles in the (100 Mpc)$^3$ simulations described
in \S~2.  As a test of the robustness of our numerical results, the
solid squares in the bottom panels of Figures~1 and 2 compare $\xi$
computed from different simulations of the CDM and the $\om=0.3$ LCDM
models.  Both simulations are performed in a much larger box -- (640
Mpc)$^3$ in comoving volume -- and a force softening distance of 160
kpc (i.e., $80\,h^{-1}$ kpc for CDM and $120\,h^{-1}$ kpc for LCDM).
The CDM simulation uses 16.8 million particles while the LCDM has 2.1
million.  As expected, the squares fall slightly below the circles at
small separation due to the larger force softening in the big-box
simulations, whereas the circles fall below the squares at comoving
separation $x\go 10\,h^{-1}$ Mpc due to their smaller box size.  In
the intermediate regime, however, the results agree well between the
two runs that have very different particle numbers, box sizes, and
force resolution.  This demonstrates that our determination of $\xi$
is robust within the resolution range of the simulations.

As discussed earlier, the nonlinear $\xi$ can also be computed from
equation~(\ref{xi}) if the corresponding nonlinear power spectrum is
known.  The solid curves in Figures~1 to 3 show the results for all
five models when the analytical approximation of Ma (1998) for the
nonlinear $P(k)$ is used as input in equation~(\ref{xi}).  The results
agree well with the nonlinear $\xi$ computed directly from the
simulation particles, providing a consistency check.

As Figures~1-3 illustrate, the effects of nonlinear gravitational
clustering drastically change the shape of the linear $\xi$ below
about $5\,h^{-1}$ comoving Mpc.  At separations of $\sim 0.1\,h^{-1}$
comoving Mpc, the nonlinear contribution boosts the linear correlation
amplitude by more than one order of magnitude.  As a result, the
nonlinear effect on the shape of $\xi$ is to straighten up the linear
curves partially, but as the solid curves show, the nonlinear $\xi(r)$
is far from being a simple power law.  In particular, $\xi$ acquires a
steeper slope in the intermediate scale around $1\,h^{-1}$ comoving
Mpc in all five models.  Such a change of slope is consistent with
similar behavior in the nonlinear power spectrum in Hamilton et
al. (1991), Jain et al. (1995), and Ma (1998).  It is also reported in
scale-free and CDM models by Padmanabhan (1996), Jain (1997), and
Jenkins et al. (1998).  This non-power-law behavior of $\xi$ of the
density field, however, should not be viewed as an inconsistency with
observations.  The $\xi$ computed from halos discussed in the next
section bears closer reality to the observed $\xi$ for galaxies, and
as will be shown, the shape of halo-halo $\xi$ is indeed well
approximated by a power law.

In the intermediate range between 1 and $10\,h^{-1}$ comoving Mpc,
Figures~1 to 3 show the interesting feature that the nonlinear $\xi$
dips below the linear $\xi$ before rising rapidly above it on smaller
scales.  The decrement is present only at $z \lo 1$, but it becomes
more prominent and extends over a larger range of $r$ as time goes on.
This feature is not due to numerical artifacts in the simulations
(such as small box size or statistical fluke) because it is seen in
all models, in both 100 Mpc and 640 Mpc simulations with different
resolution and realization, and in both $\xi$ computed from the
particles (i.e. circles and squares) and that computed from
integration of the analytical nonlinear $P(k)$ (i.e. solid curves).
Mathematically, despite the fact that the nonlinear $P(k)$ is always
larger than or equal to the linear $P(k)$ (within the $<10$\% fitting
error of Ma (1998)), the nonlinear $\xi$ can fall below the linear
$\xi$ for a limited range of $r$ because the factor $\sin(kr)/kr$ in
equation~(\ref{xi}) is oscillatory.  Physically, the decrement may
reflect the depletion of matter on quasi-linear scales that is fueling
the process of gravitational infalls in more clustered regions on
smaller scales.  The decrement also implies that the nonlinear
correlation lengths evolve more slowly than the linear predictions,
which contributes to the slower evolution of the halo-halo $\xi$ to be
discussed in the next section.  This interesting feature may warrant
further analytical study.

The dependence of $\xi(r)$ on the expansion factor $a$ is related to
the growth rate of $P(k)$, a well understood phenomenon in the linear
regime.  For the standard CDM model, the linear $P$ and $\xi$
both grow as $a^2$.  For C+HDM models, the linear $P$ and $\xi$
also increase as $a^2$ above the neutrino free-streaming scale, but
the growth is retarded below this scale because the phase-mixing of
neutrinos reduces the density perturbations.  An accurate analytic
approximation (with $<10$\% error) for this scale-dependent effect is
given by equation~(11) of Ma (1996).  For critically-flat LCDM models,
the linear $P$ and $\xi$ again grow as $a^2$ until the Universe
becomes $\Lambda$-dominated at $1+z\approx \om^{-1/3}$.  After this
epoch, the growth is slowed down (Heath 1977).  However, unlike the
C+HDM models in which the neutrino free-streaming imprints a
characteristic scale, the retardation of the linear growth rate in the
LCDM model is independent of length scales.  In the nonlinear regime,
the evolution of $\xi$ for the density field in the CDM model has been
discussed in Jain (1997) and Matarrese et al. (1997).

\subsection{Halo-Halo $\xi(r)$}
In order to determine the correlations of the simulated halos, the
Denmax algorithm (Gelb \& Bertschinger 1994ab; Ma \& Bertschinger 1994b;
Frederic 1995ab) is first used to identify dark matter halos in the
output of the simulations.  The Denmax algorithm is designed to
associate particles with density peaks by moving the particles along
the gradient of the mass density field until they settle into the
peaks.  The density field is first computed on a very fine grid from
the particle positions using the triangular-shaped cloud interpolation
scheme (Hockney \& Eastwood 1988), and then convolved with a Gaussian
filter to remove grid effects.  After the particles are associated
with density peaks by Denmax, the energy of the particles in each halo
is computed and the unbound ones are removed.  An overdensity of 200
is also imposed to select virialized halos that are potential galaxy
hosts.  Due to the small smoothing length required for identifying
galactic-size halos (typically 1/1000th the box length), this method
is more CPU-intensive than the friends-of-friends (FOF) algorithm used
in previous numerical studies of halo-halo correlation functions
(e.g., Brainerd \& Villumsen 1994; Colin et al. 1997).  The Denmax
algorithm is chosen here, however, because extensive tests have
demonstrated that it generally identifies distinct density
concentrations more reliably than the FOF algorithm, for which an
arbitrary linking length must be assigned (Gelb \& Bertschinger 1994ab;
Frederic 1995ab).

All dissipationless simulations suffer to some degree the overmerging
problem in regions of high overdensities.  When individual dark matter
halos merge in these simulations, they form a larger, smooth halo with
little substructure due to the lack of dissipation.  The extent of
this problem can be quantified by examining output of hydrodynamic
simulations, which are designed to follow both the dark matter and a
gas component that is allowed to cool and condense to high densities.
Massive dark matter halos of the size of galaxy clusters in these
simulations have been seen to host multiple subclumps of condensed gas
-- presumably the sites of galaxies -- that had survived the merger
due to their high densities and small cross sections (e.g., Katz et
al. 1992; Evrard et al. 1994).  It is not easy to identify galactic
halos in such highly clustered regions from the dark matter
distribution alone, and various methods have been proposed to locate
galactic sites in clusters in $N$-body simulations.  However, tests
against hydrodynamical simulations have so far indicated that no
method proves satisfactory on cluster scales (Summers et al. 1995).

Outside of the dense cluster environment, on the other hand, dark
matter halos identified by reliable algorithms in $N$-body simulations
have been found to give robust predictions for galaxy clustering
statistics.  For the two-point correlation function, for example,
Summers et al. (1995) found good agreement between $\xi$ determined
from dark matter halos and from the gas in hydrodynamical simulations
on scales above 1 Mpc, and most halo identification methods gave
consistent results (see their Fig. 12).  Gelb \& Bertschinger (1994b)
studied two break-up methods that were designed to reconstruct the
galactic substructure in clusters by splitting the massive halos in
$N$-body simulations into multiple sub-halos.  The amplitude of the
two-point correlation was indeed boosted significantly on scales below
1 Mpc due to the break-up procedure.  On scales above 1 Mpc, however,
both the shape and the correlation length were hardly changed whether
the massive halos were split up or not (see their Figs. 9 and 10).
These studies all indicate that the overmerging problem in $N$-body
simulations is confined to scales of $\sim 1$ comoving Mpc and below;
it therefore does not affect the determination of the correlation
length (discussed in the next section).  Moreover, most measurements
of $\xi$ are based on surveys of galaxies in blank fields devoid of
massive clusters (e.g. Shepherd et al. 1997; Le Fevre et al. 1996;
Carlberg et al. 1997).  The problem with galaxy identification in the
cluster environment therefore should not be of serious concern to this
study.

Figures~4--6 show the resulting two-point correlation functions of the
resolved halos determined from our simulations.  Five models -- the
standard CDM (Fig.~4), the $\om=0.3$ and 0.5 LCDM models (Fig.~5), and
the $\onu=0.1$ and 0.2 C+HDM (Fig.~6) -- are shown for comparison.
For each model, the figures illustrate the redshift evolution by
plotting $\xi$ from $z=0$ up to 4.  The left and right panels are
computed from virialized halos in two mass ranges, $M >
10^{11.5}\,M_\odot$ and $M > 10^{12.5}\,M_\odot$, respectively.  These
two different mass limits are chosen to illustrate the dependence of
$\xi$ on halo populations.  The figures show that the more massive
halos have a higher correlation amplitude and length.  Since the
observed $\xi$ is determined from field galaxies not associated with
massive clusters (as discussed above), I have tested the effects of
massive halos on the determination of $\xi$ by imposing an upper mass
cut of $M < 10^{14}\,M_\odot$.  Due to the rarity of these high-mass
objects, $\xi$ is found to depend little on whether these objects are
included in the sample or not.  For simplicity, such an upper mass cut
will therefore not be imposed in the results reported here.

As Figures~4-6 show, the shapes of the halo-halo $\xi$ do not depend
noticeably on the models or the halo mass cuts.  In contrary to $\xi$
for the density field (Figs~1 to 3), the halo-halo $\xi$ is well
approximated by a power law with the observed slope of $\sim -1.8$
between 1 and $10\,h^{-1}$ comoving Mpc.  On scales of below $\sim
1\,h^{-1}$ comoving Mpc, $\xi$ starts to flatten out as $z$ approaches
0 in the CDM and $\om=0.5$ LCDM models that have relatively large
small-scale power.  This behavior largely reflects the over-merging
problem in dissipationless simulations discussed above.  The
flattening of $\xi$ disappears when either halo break-up procedures
are applied (e.g., Gelb \& Bertschinger 1994b) or gas particles are
used as galaxy tracers.  Since the shape of $\xi$ carries little
information about the underlying cosmogony, no attempts will be made
here to correct for the flattening of $\xi$ on small scales.  Instead,
the focus of this study is on the more interesting quantity, the
correlation length and its evolution.  They are determined by the
reliable part of $\xi$ at $r>1\,h^{-1}$ Mpc and are largely unaffected
by the overmerging problem.

\subsection{Redshift Evolution of $\xi$ and $r_0$}
Comparison of the two-point correlation function for halos in
Figures~4-6 to that for the density field in Figures~1-3 shows that
the evolution of the halo-halo $\xi$ is markedly slower than the
matter density field in all models.  Relative to the density field,
the clustering amplitude of halos is higher at $z\sim 4$, but it
invariably drops below that of the density field by $z=0$ due to the
slow evolution.  Such ``anti-biasing'' effect has been noted in
previous studies of the CDM model (Gelb \& Bertschinger 1994b; Colin
et al. 1997); the present study shows that this effect occurs in LCDM
and C+HDM models as well.  It is attributed to the continuous
disruption and mergers of halos at late times that reduce the
clustering of halos to below that of the mass.  The relationship
between the density field and the halos shows complicated dependence
on both length and time scales and on cosmological parameters.  Care
must therefore be taken not to compare $\xi$ for the density field to
the observed $\xi$ since the former does not reflect directly the
observed clustering properties of galaxies.

The evolution of $\xi$ for halos depends strongly on the cosmological
parameters.  Figure~7 shows the redshift dependence of the physical
correlation length $r_0(z)$ for the simulated halos for two mass
ranges in all five models.  The fastest growth occurs in the pure CDM
model, in which the physical correlation length $r_0$ increases by a
factor of 8 from $z=4$ to 0.  The correlation length changes more
mildly in the LCDM and C+HDM models.  For instance, $r_0$ increases by
a factor of 3 from $z=3$ to 0 in the $\onu=0.2$ model and a factor of
5.4 from $z=4$ to 0 in the $\om=0.3$ LCDM model.  When expressed in
comoving coordinates, the halo-halo $\xi$ in fact decreases with time
at higher redshifts and is seen to rise only at a later time (see also
Brainerd and Villumsen 1994; Bagla 1997).  However, one should be
careful not to interpret this behavior as actual decrease in the
physical clustering amplitude at early times: as Figures~4 to 7
clearly illustrate, the physical $r_0$ and $\xi(r)$ always increase
monotonically with time in all models.  Instead, the decrease merely
reflects the fact that the halo clustering rate is slower than the
state of fixed clustering in comoving coordinates early on.  It
eventually surpasses fixed comoving clustering when the comoving
correlation length begins to increase (see Fig.~8 below).

The top panels of Figure~7 illustrate that $r_0$ increases more slowly
in models with smaller $\om$.  This feature is consistent with the
slower growth of structure in low-density models.  The lower panels
compare different C+HDM models and indicate that larger $\onu$ models
show slower evolution.  This can be explained by the slower growth due
to the large neutrino free-streaming effects in high-$\onu$ models.
The left and right panels of Figure~7 illustrate that $r_0$ also
depends on the minimal halo mass selected for the sample, and is
larger for the more massive halos (right panels) due to their high
clustering amplitudes.

To quantify the evolution of the two-point correlation function with
redshift, I find it most convenient to parameterize the physical 
and comoving correlation lengths directly as
\begin{eqnarray}
	r_0(z) &=& (1+z)^{-\eta(z)}\, r_0(0)\,, \nonumber\\
	x_0(z) &=& (1+z)^{-\eta(z)+1}\, x_0(0)\,.
\label{r0x0}
\end{eqnarray}
The parameter $\eta$ itself increases with time because as Figure~7
illustrates, the slope of $r_0(z)$ generally steepens as $z$
decreases, and the rate of change depends on $\om$ and $\onu$.  The
form of equation~(\ref{r0x0}) is chosen to improve the power-law
parameterization commonly used in the literature, where
both the shape and redshift dependence of $\xi$ are
written as power laws with exponents $\gamma$ and $\epsilon$:
\begin{eqnarray}
	\xi(r,z)&=&\left( {r\over r_0(z)}
	\right)^{-\gamma}\,,\nonumber \\
	\xi(r,z)&=&(1+z)^{-3-\epsilon} \xi(r,0)\,,
\label{xir}
\end{eqnarray}
and $r$ is the pair separation in physical coordinates.  It follows
from equation~(\ref{xir}) that the physical and comoving correlation
length evolve as
\begin{eqnarray}
	r_0(z) &=& (1+z)^{-3-\epsilon \over \gamma} r_0(0)\,,\nonumber\\
    x_0(z) &=& (1+z)^{-3-\epsilon+\gamma \over \gamma} x_0(0)\,.
\label{r0}
\end{eqnarray}
As Figure~7 illustrates, however, the power-law redshift dependence in
equation~(\ref{r0}) does not hold over the redshift range 0 to 4 for
any model.  Moreover, since the parameter $\epsilon$ in
equation~(\ref{xir}) was introduced to characterize the redshift
evolution of $\xi$ and not $r_0$, the evolution of $r_0$ depends
inconveniently on both $\epsilon$ and $\gamma$.  The simple form of
equation~(\ref{r0x0}) will therefore be used here.  To translate
$\eta$ into the parameter $\epsilon$ used in the literature, use
\begin{equation}
	\epsilon={\gamma\,\eta-3}\,.
\end{equation}

Some useful theoretical predictions for the evolution of $\xi$ are:
(1) The linear theory predicts $\xi\propto (1+z)^{-2}$ for the
standard CDM model, corresponding to $\epsilon=\gamma-1$, or
$\epsilon\approx 0.8$ and $\eta\approx 2.1$ for $\gamma\approx 1.8$;
(2) clustering fixed in comoving coordinates corresponds to
$x_0(z)=x_0(0)$, implying $\eta=1$, or $\epsilon\approx -1.2$; (3)
stable clustering in physical space corresponds to $\xi\propto
(1+z)^{-3}$, and therefore $\epsilon=0$, or $\eta\approx 1.7$.

Figure~8 shows $\eta(z)$ for the simulated halos in our study.  The
rise of $\eta(z)$ with time indicates that the halo correlation length
$r_0$ grows faster than a simple power law with a constant $\epsilon$
assumed in the literature (as seen in Figure~7).  The parameter
$\eta(z)$ also depends on both cosmological parameters $\om$ and
$\onu$.  The label on the right along the vertical axis marks the
corresponding $\epsilon$ when $\gamma=1.8$ is assumed for the slope of
$\xi$.  When $\eta$ crosses unity in a given model, it indicates that
gravitational clustering has become faster than no evolution in
comoving space (i.e., $x_0(z)=x_0(0)$).  This epoch corresponds to the
turn-around redshift at which the initially falling $\xi$ first starts
to increase in comoving space.  At $z\approx 0$, the CDM, $\om=0.5$
LCDM, and $\onu=0.1$ C+HDM models all approach the stable clustering
regime of $\eta\approx 1.7$, whereas the $\om=0.3$ LCDM model exhibits
less rapid evolution of $\eta\approx 1.3$.  Stable clustering has also
been seen for the density field on small scales in the pure CDM model
and models with scale-free spectra (Jain 1997).  Our study here shows
that halos also approach stable clustering in high-$\om$ models.

The only detailed numerical study of the evolution of halo-halo $\xi$
thus far is Colin et al. (1997), in which the $\om=1$ and $\om=0.2$
CDM models are examined.  But as discussed in \S 1, the $\om=1$ CDM
results reported here cannot be compared directly to theirs because of
their choice of high Hubble constant $h=1$ and large value of the
shape parameter $\Gamma=1$.  Since the standard CDM model with
$\Gamma=0.5$ used in this paper already suffers from the well-known
excess small-scale power, a model with $\Gamma=1$ would produce even
more power and worsen the problem.  This may be the main cause of the
steep slope $\gamma \approx 2.2$ to 2.4 for $\xi$ in Table 2 of Colin
et al., which is significantly larger than the value of $\gamma
\approx 1.8$ found here and in other works.  One should also note that
because $r_0(z)$ depends on both $\gamma$ and $\epsilon$ (see
eq.~[\ref{r0}]), their high values of $\gamma$ lead to the large
positive $\epsilon$ in their Figure~9 (where $\epsilon$ ranges from
$\sim 0$ at $z=5$ to 1.5 at $z=1$). When expressed in terms of the
parameter $\eta$ introduced earlier, their large values of $\gamma$
and $\epsilon$ do cancel out to some extent, giving $\eta\approx 1.3$
at $z=5$ and 1.87 at $z=0$, but it is still significantly larger than
our $\epsilon$ in Figure~8.

\section{Conclusions and Discussion}
Motivated by the extensive observational effort toward determining the
evolution of the two-point correlation function $\xi(r)$, I have
presented a detailed theoretical study of the the redshift dependence
of $\xi$ in five cosmological models with varying mass density $\om$
and neutrino fraction $\onu$.  The evolution is found to be much more
intricate than the simple power-law parameterization of $\epsilon$
commonly assumed in the literature, and a redshift-dependent variable
$\eta$ is introduced to quantify the evolution of the correlation
length: $r_0(z)=(1+z)^{-\eta(z)}\,r_0(0)$.  The value of $\eta(z)$
depends strongly on $\om$ and $\onu$, and is smaller in models with
lower $\om$ or higher $\onu$ (Fig.~8), indicating a slower growth in
$r_0$.  Within a given model, $\eta$ increases with time, passing the
phase of fixed comoving clustering ($\eta=1$) at $z\sim 1$ to 3 toward
stable clustering ($\eta\approx 1.7$) at $z\sim 0$.

Although the two-point correlation function $\xi$ provides a
fundamental statistical measure of gravitational clustering, care must
be taken in specifying the underlying quantity that is being measured
and in including the possible nonlinear effects.  This study has
demonstrated that $\xi$ for the matter density field and dark matter
halos have different shapes and evolve with different rates.  While
the local slope of $\xi$ for the nonlinear density field changes with
length scales (Figs.~1 to 3), the shape of the halo-halo $\xi$ is well
approximated by a power law with the observed slope of $\sim -1.8$
independent of cosmological parameters (Figs.~4 to 6).  It is
therefore the redshift evolution of the amplitude of $\xi$ that will
reveal the most about cosmogonies.  Since halos generally correspond
to galaxy sites (except in dense cluster environment), $\xi$ for halos
in simulations is a reasonable quantity to compare with observations.
The correlation function for the density field provides important
insight for theoretical studies of gravitational clustering, but it
bears little direct relevance to the observed clustering of galaxies.

On the observational side, ongoing deep redshift surveys of galaxies
that cover larger areas of the sky should soon yield more robust
determination of $\xi$ beyond $z\approx 0$.  Photometric redshifts
determined from multi-color surveys and the two-point angular
correlation function also offer complimentary measurements of $\xi$
(Connolly, Szalay, \& Brunner 1998; Postman et al. 1998; and
references therein).  We are currently analyzing galaxy clustering
properties in two redshift surveys conducted with the Norris
Spectrograph on the Palomar 200-inch telescope (Small et al. 1998).
The combined surveys have more than 1000 galaxies in $0.2 < z < 0.5$.
The implications of the theoretical results presented here will be
discussed in Small et al. (1998).

Although the evolution of $\xi$ has already been counted as yet
another strike against $\om=1$ models (Carlberg et al. 1997), it must
be borne in mind that the nature of high-redshift galaxies has only
begun to be unravelled, and that little is known about their dark
matter halos.  Due to the large uncertainties in the current
observational results and theoretical understanding in galaxy
evolution, this paper has chosen to investigate the evolution of $\xi$
using simulated dark matter halos within well defined mass ranges.  It
is nonetheless conceivable that measurements of $\xi$ at higher
redshifts are probing different galaxy populations that are not fair
counterparts of the local galaxies.  Based on future observations, one
should be able to refine the theoretical predictions made in this
paper by incorporating different dark matter halo populations at
higher redshifts.

\acknowledgments 
The author thanks Jim Frederic and Todd Small for valuable
discussions.  Supercomputing time was provided by the National Scalable
Cluster Project at the University of Pennsylvania and the National
Center for Supercomputing Applications.

\clearpage
%\section*{Figure Captions}
%% Fig 1
\begin{figure}
\epsfxsize=6.5truein 
\epsfbox{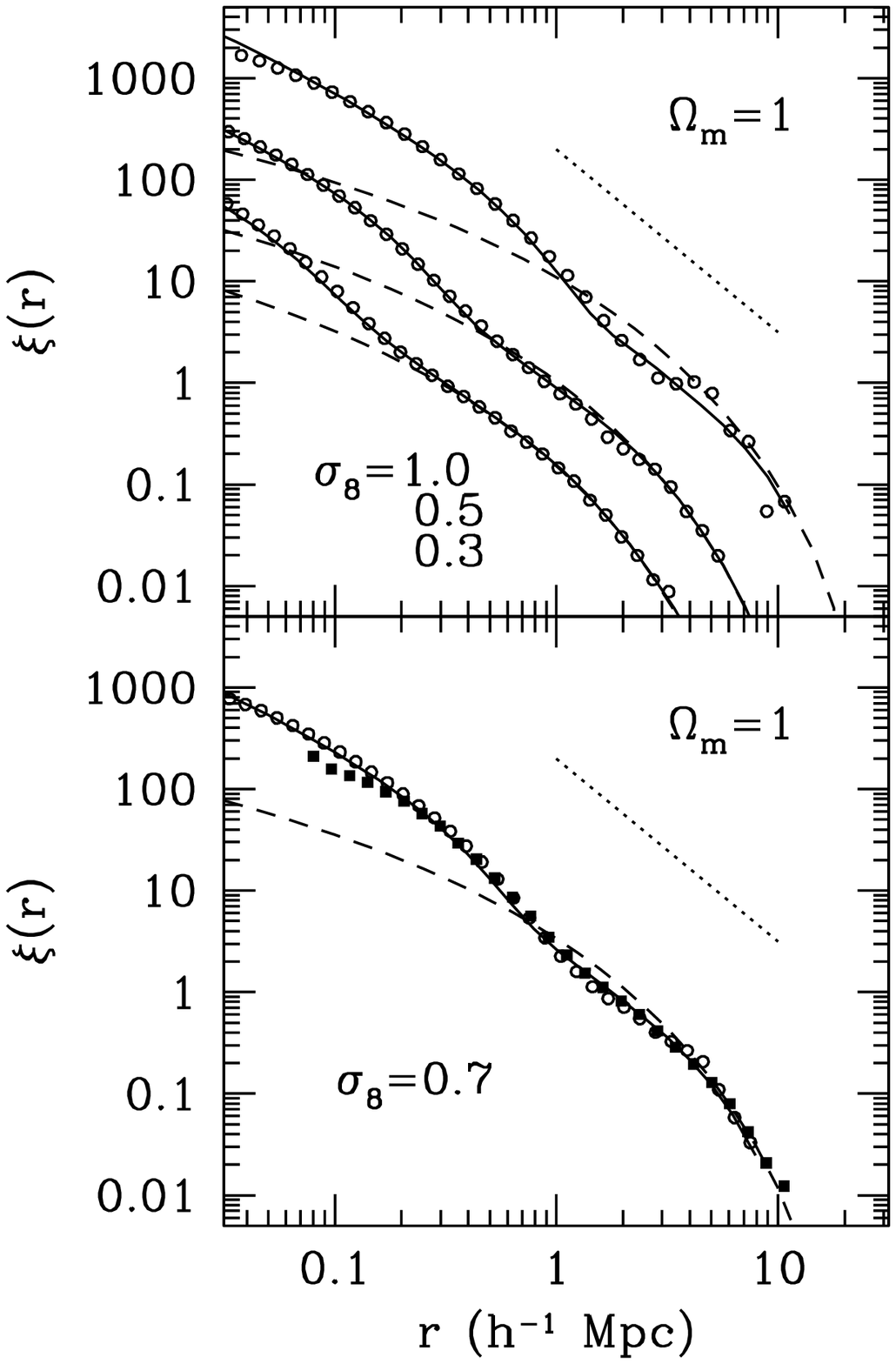}
\caption{Evolution of the two-point correlation function $\xi(r)$ (in
physical coordinates) for the matter density field in the standard CDM
model with $\om=1$ and $h=0.5$.  The top panel shows results at
$\sigma_8=0.3$, 0.5, and 1.0 (from bottom up), and the bottom panel
shows $\sigma_8=0.7$ separately for clarity.  (The COBE normalization
corresponds to $\sigma_8=1.4$ at $z=0$, while $\sigma_8\approx 0.7$ is
required to match the observed abundance of $z\approx 0$ galaxy
clusters.)  The dashed curves show linear-theory predictions.  The
symbols represent the nonlinear $\xi$ computed from particles in the
simulations -- the circles are from a simulation in a (100 Mpc)$^3$
box, and the solid squares for the $\sigma_8=0.7$ output is from a
(640 Mpc)$^3$ simulation.  The solid curves, which agree well with the
symbols, represent the nonlinear $\xi$ from integration of the
analytic nonlinear $P(k)$ in Ma (1998).  The dotted straight line
indicates a slope of $-1.8$ in this and the following figures.}
\end{figure}

%% Fig 2
\begin{figure}
\epsfxsize=6.5truein 
\epsfbox{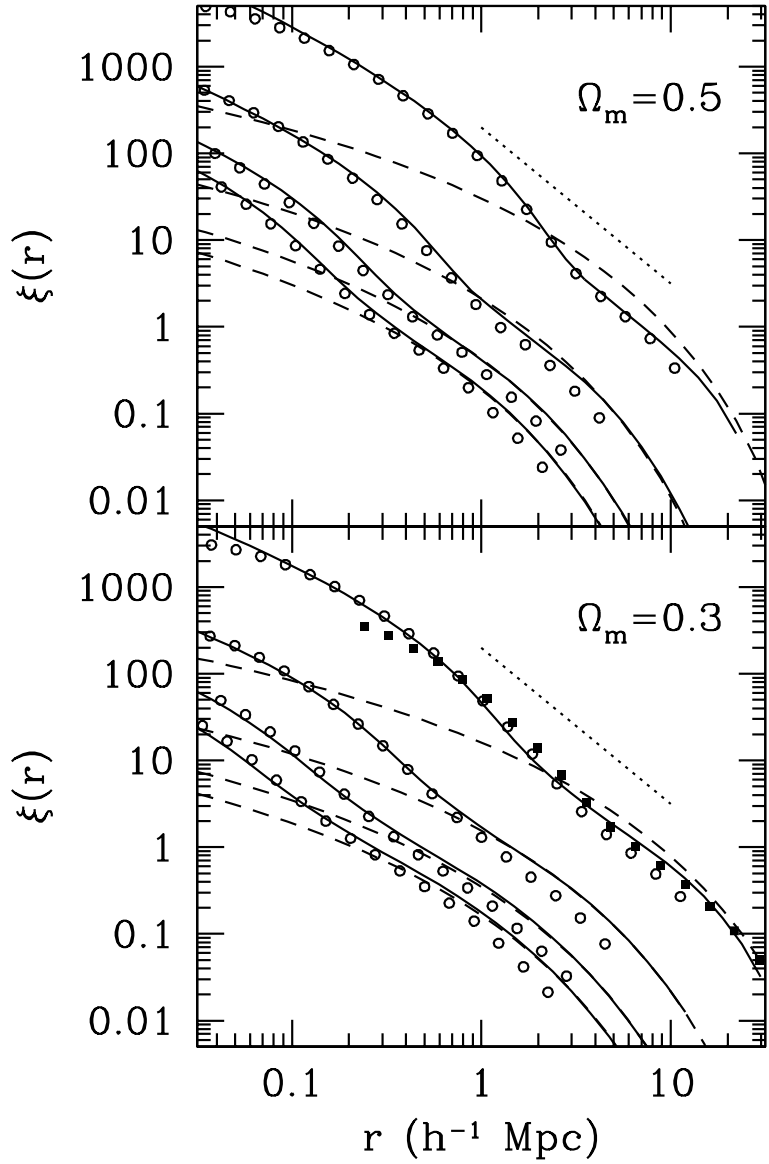}
\caption{Same as Fig.~1 but for two LCDM models: $\om=0.5$, $\ov=0.5$,
and $h=0.7$ (top); and $\om=0.3$, $\ov=0.7$, and $h=0.75$ (bottom).
For each model, four redshifts are shown: $z=4$, 3, 1.5, and 0 (from
bottom up).  The solid squares compare $\xi$ computed from a large (640
Mpc)$^3$ simulation of the $\om=0.3$ model.}
\end{figure}

%% Fig 3
\begin{figure}
\epsfxsize=6.5truein 
\epsfbox{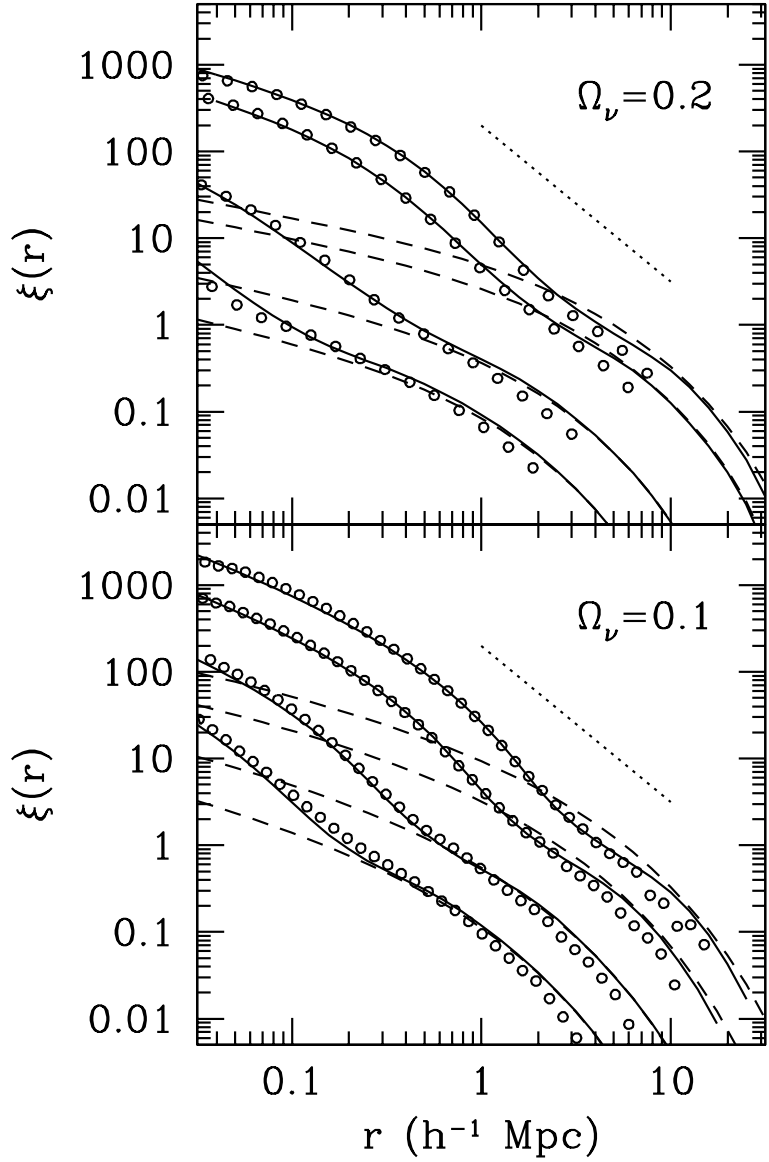}
\caption{Same as Fig.~1 but for two C+HDM models: $\onu=0.2$,
$\oc=0.75$ (top); $\onu=0.1$, $\oc=0.85$ (bottom).  Both have
$\ob=0.05$ and $h=0.5$.  The redshifts shown are (from bottom up):
$z=3$, 1.5, 0.26 and 0 for $\onu=0.2$, and $z=3$, 1.5, 0.43 and 0 for
$\onu=0.1$.}
\end{figure}

%% Fig 4
\begin{figure}
\epsfxsize=6.5truein 
\epsfbox{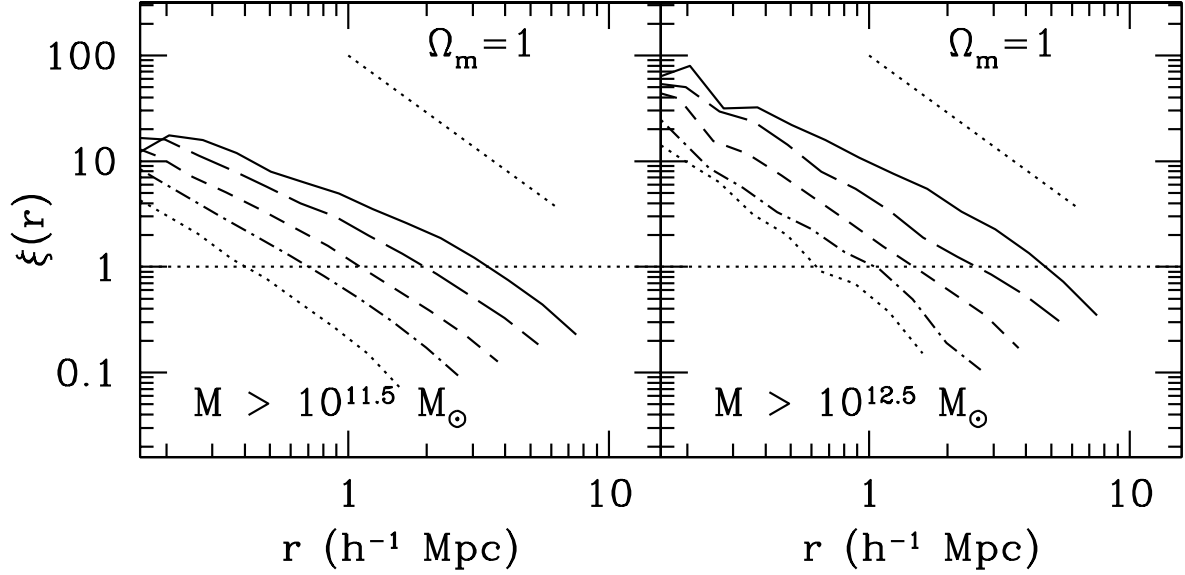}
\caption{Evolution of the two-point correlation function $\xi(r)$ for
the simulated dark matter halos in the standard CDM model with $\om=1$
and $h=0.5$.  Five epochs are shown: $\sigma_8=0.3$ (dotted), 0.5
(dot-short-dashed), 0.7 (short-dashed), 1.0 (long-dashed), and 1.4
(solid).  (The COBE normalization corresponds to $\sigma_8=1.4$ at
$z=0$.)  The left and right panels show $\xi$ computed from virialized
halos in the mass range $M > 10^{11.5}\,M_\odot$ and $M >
10^{12.5}\,M_\odot$, respectively.  The dotted straight line indicates
the observed slope of $-1.8$.}
\end{figure}

%% Fig 5
\begin{figure}
\epsfxsize=6.5truein 
\epsfbox{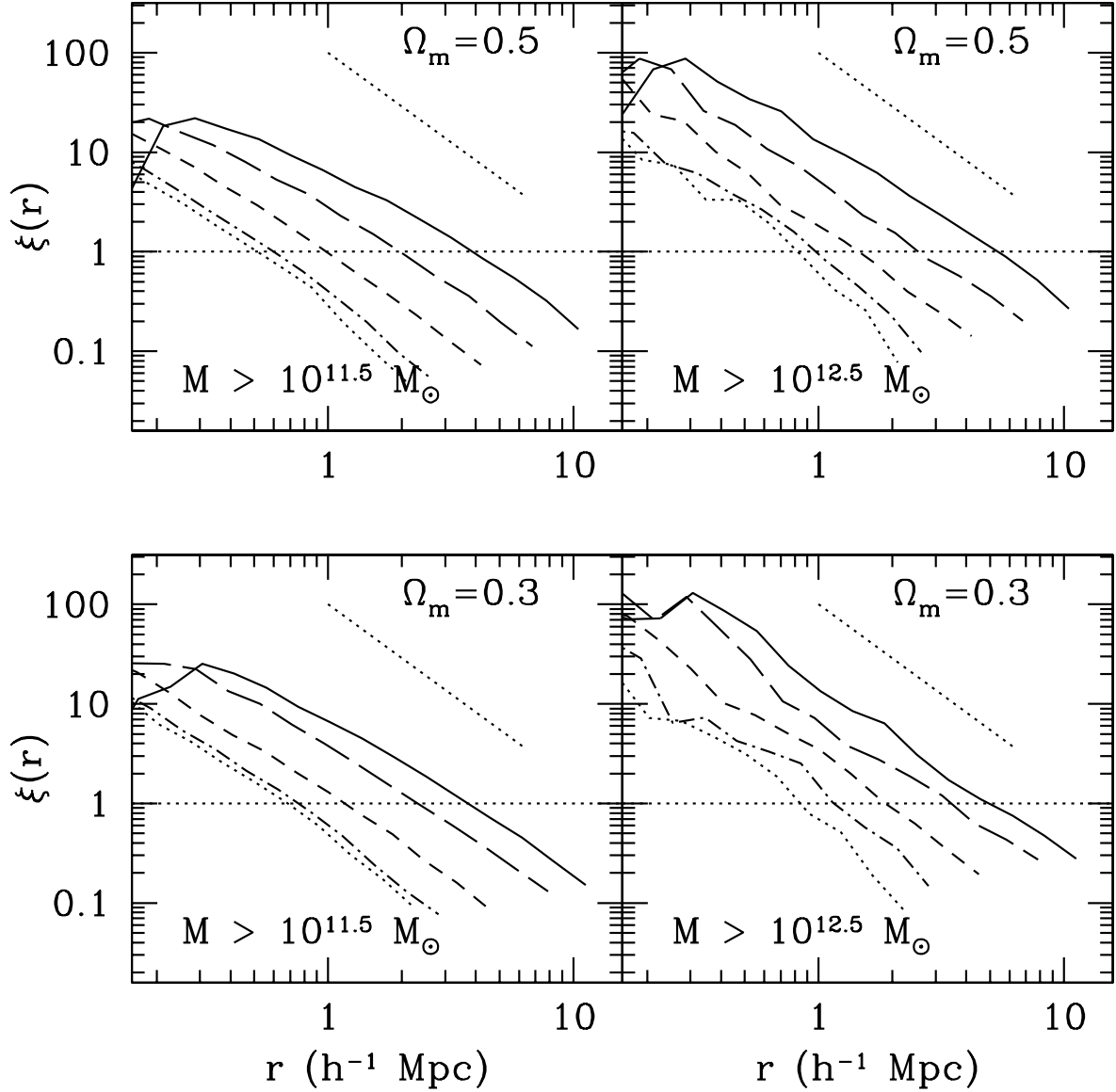}
\caption{Same as Fig.~4 but for the two LCDM models: $\om=0.5$,
$\ov=0.5$, and $h=0.7$ (top); and $\om=0.3$, $\ov=0.7$, and $h=0.75$
(bottom).  Five epochs are shown: $z=4$ (dotted), 3
(dot-short-dashed), 1.5 (short-dashed), 0.5 (long-dashed), and 0
(solid).}
\end{figure}

%% Fig 6
\begin{figure}
\epsfxsize=6.5truein 
\epsfbox{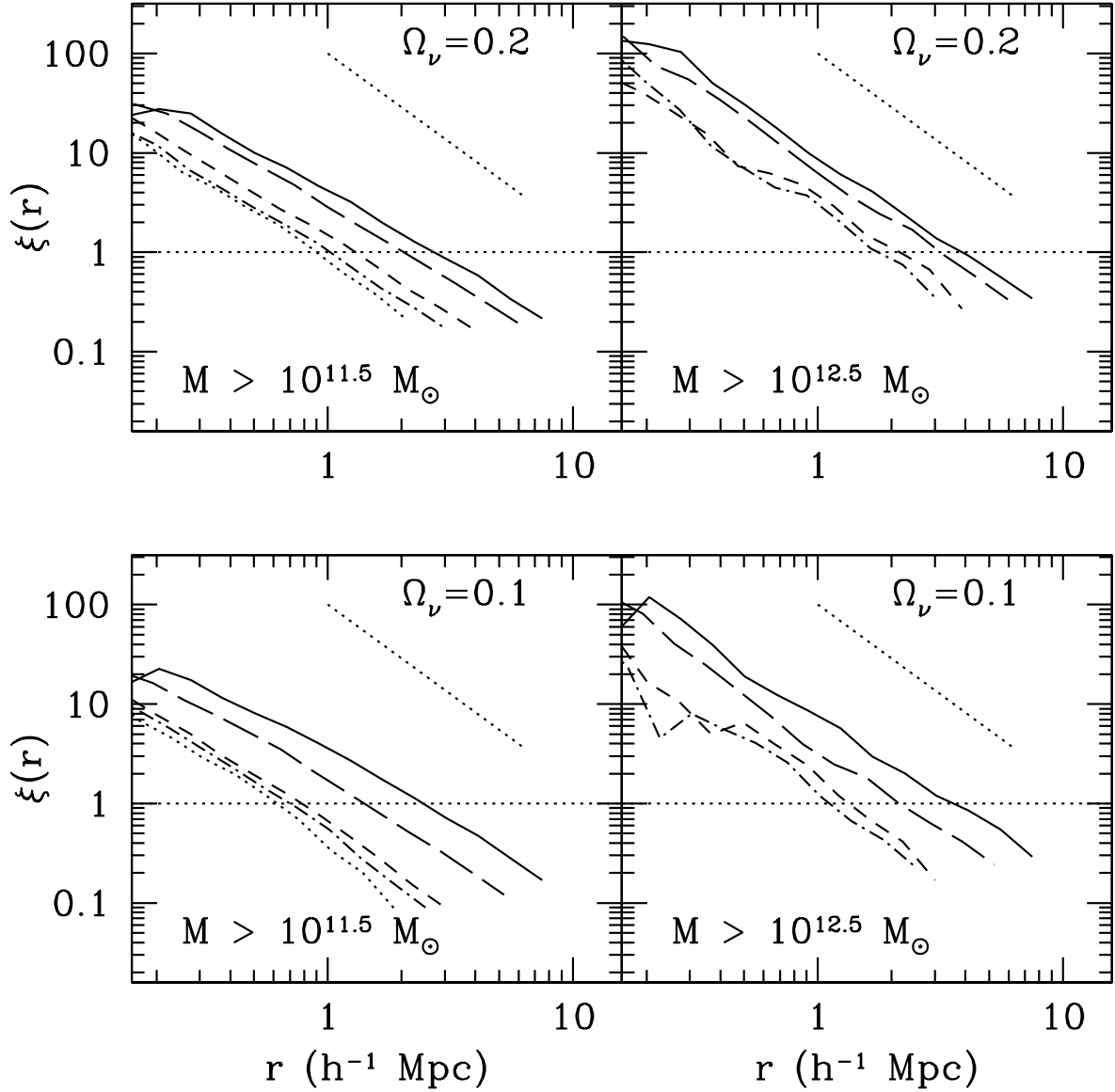}
\caption{Same as Fig.~4 but for the two C+HDM models -- $\onu=0.2$
(top) and $\onu=0.1$ (bottom).  For $\onu=0.2$, the redshifts shown
are: $z=2.6$ (dotted), 1.5 (dot-short-dashed), 1.0 (short-dashed),
0.26 (long-dashed), and 0 (solid).  For the $\onu=0.1$ model, the
redshifts are: $z=3$ (dotted), 2 (dot-short-dashed), 1.5
(short-dashed), 0.43 (long-dashed), and 0 (solid).  (The dotted curves
are omitted from the right panels since estimates of $\xi$ are very
noisy due to the small number of massive halos at such early epochs
formed in C+HDM models.)}
\end{figure}

%% Fig 7
\begin{figure}
\epsfxsize=6.5truein 
\epsfbox{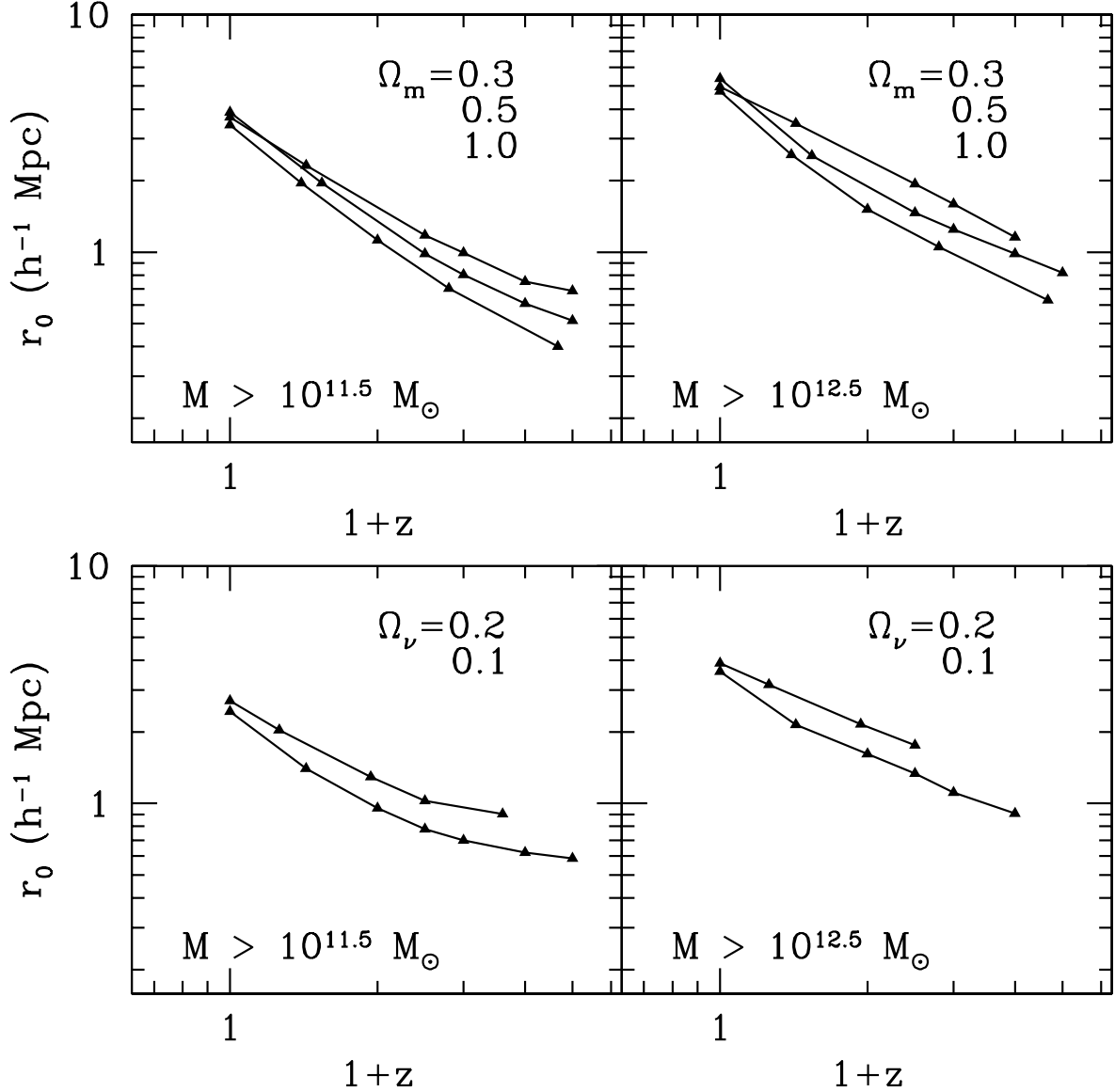}
\caption{Redshift evolution of the physical correlation length $r_0$
for the simulated halos.  In the top panels, the three curves from top
down are for the $\om=0.3$, 0.5, and 1.0 LCDM models.  In the bottom
panels, the two curves are for the $\onu=0.2$ (top) and 0.1 (bottom)
C+HDM models.  The redshift dependence of $r_0$ varies with
cosmological parameters and is not a power law (see eq.~[3]) as has
often been assumed.  The left and right panels compare two different
halo mass ranges and show that more massive halos generally have a
larger $r_0$.}
\end{figure}

%% Fig 8
\begin{figure}
\epsfxsize=6.5truein 
\epsfbox{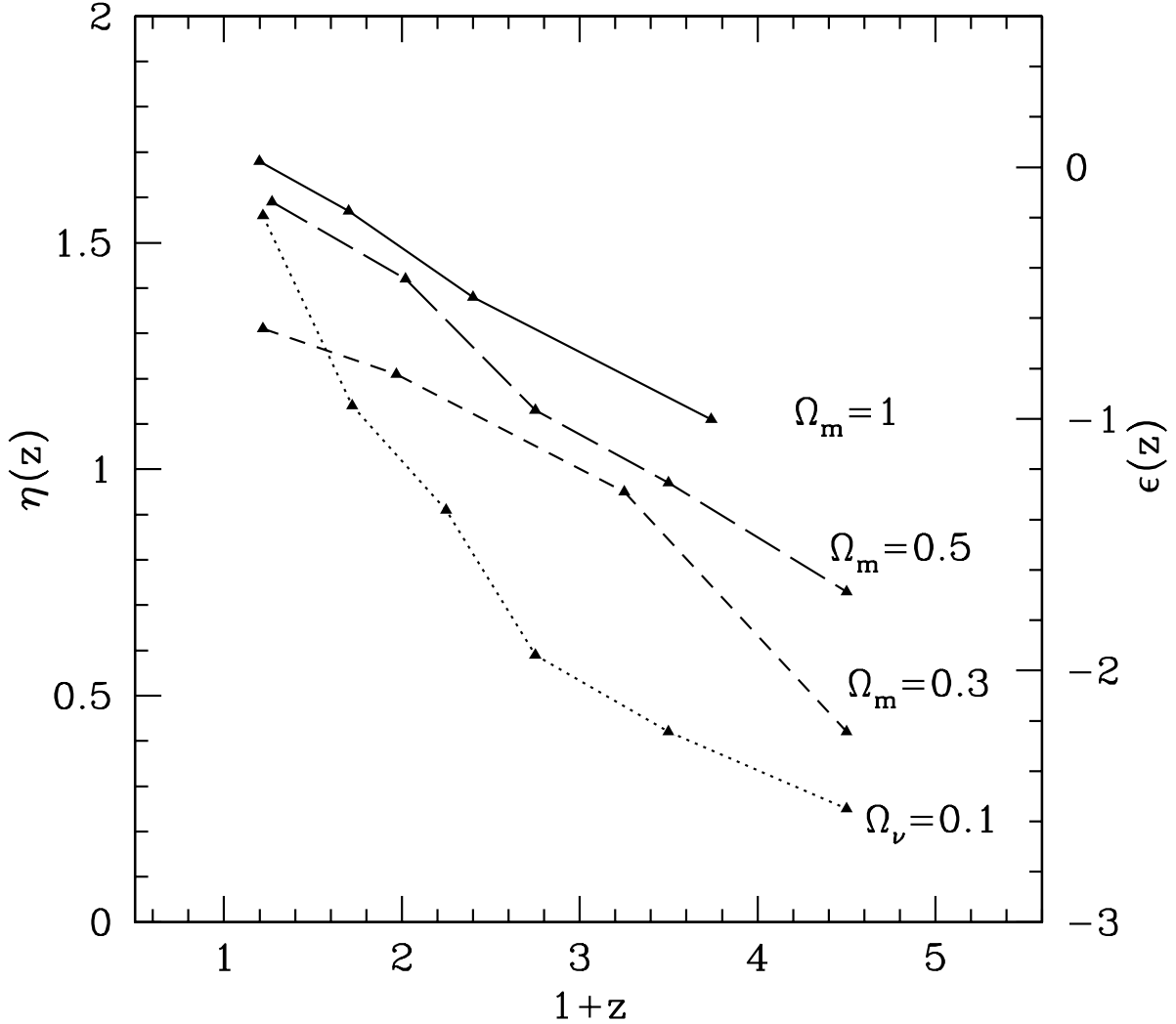}
\caption{The evolution parameter $\eta(z)$, defined as
$r_0(z)=(1+z)^{-\eta(z)}\,r_0(0)$, for the $M> 10^{11.5}\,M_\odot$
virialized halos in four models: CDM, $\om=0.5$ LCDM, $\om=0.3$ LCDM,
and $\onu=0.1$ C+HDM models.  The COBE normalization is used,
including the CDM.  The vertical labels on the right mark the
corresponding $\epsilon$ used in the literature, where slope of
$\gamma=1.8$ has been assumed to relate $\eta$ and $\epsilon$:
$\epsilon=1.8\eta-3$ (see eq.~[7]).  The rise of $\eta$ with
decreasing $z$ shows that the halo correlation length increases faster
than a simple power law assumed in the literature.  The clustering
passes $\eta=1$ (i.e. $\epsilon=-1.2$) for fixed clustering in
comoving coordinates and approaches $\eta=1.7$ ($\epsilon=0$) for
stable clustering at $z\approx 0$.}
\end{figure}

\end{document}